\documentclass{article}

\usepackage{arxiv}

\usepackage[utf8]{inputenc} 
\usepackage[T1]{fontenc}    
\usepackage{hyperref}       
\usepackage{url}            
\usepackage{booktabs}       
\usepackage{amsfonts}       
\usepackage{nicefrac}       
\usepackage{microtype}      
\usepackage{lipsum}		
\usepackage{graphicx}
\usepackage{doi}
\usepackage{multirow}
\usepackage{amsmath}
\usepackage{algorithm}
\usepackage[numbers]{natbib}
\usepackage{multirow}

\title{Finger-to-Chest Style Transfer–assisted Deep Learning Method For  Photoplethysmogram Waveform Restoration with Timing Preservation}

\author{ \href{https://orcid.org/0000-0001-9204-3363}{\includegraphics[scale=0.06]{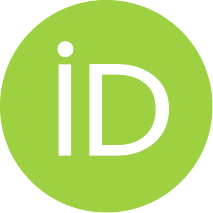}\hspace{1mm}Sara Maria Pagotto} \\
	Department of Electronics,\\
	Information, and Bioengineering,\\
	Politecnico di Milano, Milan, Italy. \\
     \And
 {\includegraphics[scale=0.06]{orcid.pdf}\hspace{0.1mm}Federico Tognoni} \\
	Department of Electronics,\\
	Information, and Bioengineering,\\
	Politecnico di Milano, Milan, Italy. \\
     \And
     \href{https://orcid.org/0000-0003-2519-0720}
 {\includegraphics[scale=0.06]{orcid.pdf}\hspace{0.1mm}Matteo Rossi} \\
	Department of Electronics,\\
	Information, and Bioengineering,\\
	Politecnico di Milano, Milan, Italy. \\
    \And
	\href{https://orcid.org/0000-0002-2654-9734}
 {\includegraphics[scale=0.06]{orcid.pdf}\hspace{0.1mm}Dario Bovio} \\
	Biocubica srl, Milan, Italy. \\        
    \And
    \href{https://orcid.org/0000-0002-2944-2254}
  {\includegraphics[scale=0.06]{orcid.pdf}\hspace{0.1mm}Caterina Salito} \\
	Biocubica srl, Milan, Italy. \\ 
    \And
	\href{https://orcid.org/0000-0002-6276-6314}
 {\includegraphics[scale=0.06]{orcid.pdf}\hspace{0.1mm}Luca Mainardi} \\
	Department of Electronics,\\
	Information, and Bioengineering\\
	Politecnico di Milano, Milan, Italy. \\
    \And
	\href{https://orcid.org/0000-0003-3995-8673}
 {\includegraphics[scale=0.06]{orcid.pdf}\hspace{0.1mm}Pietro Cerveri} \\
	Department of Electronics,\\
	Information, and Bioengineering\\
	Politecnico di Milano, Milan, Italy. \\
        Università di Pavia, Pavia, Italy.
        \texttt{pietro.cerveri@unipv.it} \\
}

\date{}

\hypersetup{
pdftitle={Finger-to-Chest Style Transfer–assisted Deep Learning Method For  Photoplethysmogram Waveform Restoration with Timing Preservation},
pdfsubject={cs.CV, cs.AI},
pdfauthor={Sara Maria Pagotto, Federico Tognoni, Matteo Rossi, Dario Bovio, Caterina Salito, Luca Mainardi, Pietro Cerveri},
pdfkeywords={Photoplthysmogram, Deep Learning, Style Transfer, Signal Restoration},
}

\begin{document}
\maketitle

\begin{abstract}
Wearable measurements, such as those obtained by  photoplethysmogram (PPG) sensors are highly susceptible to motion artifacts and noise, affecting cardiovascular measures. Chest-acquired PPG signals are especially vulnerable, with signal degradation primarily resulting from lower perfusion, breathing-induced motion, and mechanical interference from chest movements. Traditional restoration methods often degrade the signal, and supervised deep learning (DL) struggles with random and systematic distortions, requiring very large datasets for successful training. To efficiently restore chest PPG waveform, we propose a style transfer–assisted cycle-consistent generative adversarial network, called starGAN, whose performance is evaluated on a three-channel PPG signal (red, green, and infrared) acquired by a chest-worn multi-modal sensor, called Soundi$^\text{\textregistered}$. Two identical devices are adopted, one sensor to collect the PPG signal on the chest, considered to feature low quality and undergoing restoration, and another sensor to obtain a high-quality PPG signal measured on the finger, considered the reference signal.  Extensive validation over some 8,000 5-second chunks collected from 40 subjects showed about 90\% correlation of the restored chest PPG with the reference finger PPG, with a 30\% improvement over raw chest PPG. Likewise, the signal-to-noise ratio improved on average of about 125\%, over the three channels. The agreement with heart-rate computed from concurrent ECG was extremely high, overcoming 84\% on average. These results demonstrate effective signal restoration, comparable with findings in recent literature papers. Significance: PPG signals collected from wearable devices are highly susceptible to artifacts, making innovative AI-based techniques fundamental towards holistic health assessments in a single device.
\end{abstract}


\section{Introduction}
\label{sec:introduction}
Physiological signals acquired from wearable devices, such as the photoplethysmogram (PPG), face ongoing challenges due to substantial motion artifacts and noise, thus demanding waveform denoising and restoration \cite{GarciaLopez2020, Marzorati2022}. It has been extensively documented in the literature that ineffective procedures may often induce modifications or even the degradation of the underlying cardiovascular information extracted from the PPG \cite{Dao2017}, making wearable devices less reliable. High-quality signals are essential to broaden the application of PPG in diverse cardiac features, including stress detection, mental workload assessment, emotion classification, and sleep quality monitoring \cite{Rossi2023, Fan2024, Gupta2024}. For example, heart rate variability (HRV), which could be extracted from the PPG signal, is a valuable predictor of many cardiac conditions and arrhythmic events. The spectral components of HRV also reveal valuable information about parasympathetic and sympathetic nervous system activity. Nonetheless, PPG is currently subject of intensive research for its role in the arterial blood pressure computation \cite{Marzorati2020,Wang2024}.  
All such physiological relevant features are primarily obtained from high-quality pulse waveforms measured by contact-based PPG methods. Therefore, advancing PPG technology to accurately extract pulse waveforms is crucial for calculating a broader range of physiological metrics. 
Current trends in the biosignal processing, including the PPG, focus on advanced data-driven artificial intelligence tools, as end-to-end deep learning (DL) networks \cite{Rossi2022, Mohagheghian2024}. They automatically encode the significant feature describing the intrinsic nature of the physiological waveforms and exploit that to restore the signal quality in the presence of artifacts using a decoder. Nonetheless, such methods often experience difficulties in denoising the inherent bodily signals due to many random and systematic in-band distortions. AI-based techniques typically require large datasets to effectively estimate model parameters, minimizing the risks of overfitting and bias. Furthermore, supervised deep learning (DL) methods require a precise alignment between input and reference signals in the training dataset. This requirement presents a challenge for photoplethysmogram (PPG) restoration, as it can be difficult to simultaneously obtain both high-quality reference signals and low-quality input signals. A common workaround involves generating simulated corrupted signals by introducing noise to high-quality acquired signals. However, simulated noise is hardly equivalent to real artifacts.
To address these research challenges, this study developed a novel cycle-consistent generative adversarial network (cycle-GAN) architecture, termed starGAN, for restoring 3-channel PPG waveforms acquired at the chest. The restoration process utilized simultaneously acquired PPG signals from the finger as a reference, to transfer the finger PPG style to the chest PPG while preserving the original timing. Data acquisition was performed using Soundi sensor (Biocubica srl, Milan, Italy). The experiments involved two identical, synchronized Soundi sensors: the first collected chest PPG signals requiring denoising, while the second captured the corresponding reference PPG signals from the finger.  The main innovative contribution consist of:

\begin{itemize}
    \item developing a novel Cycle-GAN architecture specifically designed for restoring photoplethysmogram (PPG) waveforms,
    \item preserving the original timing of the chest PPG signal by properly weighting the cycle and identity losses effects,    
    \item proposing an experimental setup that avoids simulation of artifacts,
    \item exploiting three-channel PPG sensor fusion to increase quality restoration;
    \item validating cardiac-related quantities computed on the restored PPG against quantities extracted from ECG recorded simultaneously.
\end{itemize}

\subsection{Related Works}
The processing and denoising of photoplethysmography (PPG) signals have been representing so far a critical area of research, driven by the growing demand for reliable wearable devices and clinical applications. Traditional filtering techniques, such as Butterworth and Bessel filters, were foundational in removing baseline drifts and high-frequency noise from PPG signals. A Wiener filter was proposed to reduce PPG signal noise \cite{Temko2017}. However, limitations were evident when confronted with non-stationary noise, such as systematic motion artifacts or sudden light interference, which are common in real-world settings, requiring extensive post-processing. Dao et al. \cite{Dao2017} advanced this domain by proposing a motion artifact detection algorithm, based on  time-frequency spectrum processing, capable of addressing such challenges, although the trade-off between artifact removal and signal distortion remained an open issue. Time-frequency analysis based on wavelet transform was proposed to denoise wrist PPG, which was proven effective on the removal of high-frequency components \cite{Gupta2023}. 
The application of machine learning (ML) and deep learning (DL) has marked a paradigm shift in PPG signal analysis. Autoencoders, as employed by Lee et al. and Gautam et al. \cite{Lee2019, Gautam2024}, were proven effective in extracting features and mitigating noise. Bidirectional recurrent autoencoders, in particular, demonstrate the ability to capture temporal dependencies, making them well-suited for time-series data like PPG. These architectures can differentiate between high quality physiological patterns and artifacts, enhancing signal clarity while retaining critical information. Hybrid convolutional neural networks (CNNs), emphasizing the integration of spatial and temporal features for end-to-end evaluation of PPG signals, were proposed \cite{Marzorati2022}, achieving high accuracy in feature detection tasks. Generative approaches further pushed the boundaries of PPG signal processing \cite{Song2021, Mahmud2024}. 
The PulseGAN framework was reported to improve PPG signal quality, enhancing heart rate and HRV accuracy as well. Nonetheless, its reliance on denoising chrominance signals and adversarial loss could limit generalization across diverse datasets \cite{Song2021}. Cycle-coherent generative adversarial networks (ccGANs) were proposed to address the lack of paired data, a common challenge in biomedical datasets \cite{Mahmud2024}. The model ccGANs exploited unpaired data to establish domain mappings, allowing noisy wrist or chest PPG signals to be transformed into higher-quality finger PPG signals. However, challenges such as GAN sensitivity, modest improvements in HR/HRV correlations, generalizability across diverse conditions, and computational complexity for real-time applications claimed further investigation.
Recently, Hu et al. \cite{Hu2024} explored hybrid denoising approaches that blend conventional filtering with ML, achieving a balance between artifact suppression and signal integrity.
Despite these advances, challenges remain. First, most of the restoration methods often struggle with severe motion artifacts and unpredictable noise in wearable biosignals, failing to generalize across various motion scenarios \cite{Park2022}. Second, few existing solutions leverage  sensor fusion, using for instance acceleration data \cite{Xu2020, Talukdar2022}, however lacking the integration of multi-channel PPG, such as red, green, and infrared  signals, to improve restoration performance. Third, traditional DL methods reported over-smoothing signals during noise removal, compromising the physiological integrity of the biosignal and negatively affecting downstream clinical metrics like heart rate or blood pressure estimation \cite{Mahmud2024}. Finally, most studies relied on publicly available datasets, using very small cohorts, and performed simulated corruption of the original  PPG signals \cite{Gupta2023, AvilaCastro2025}, which sensibly limited the span of the findings. By addressing these gaps, the proposed style transfer–assisted cycle-consistent GAN (starGAN) can significantly advance the field by enabling efficient signal restoration, leveraging multi-channel data, avoiding supervised learning, and preserving clinical integrity in wearable biosignal analysis.

\section{Methodology}
\subsection{Chest-worn device}
The acquisition device adopted in this work was the patented (no. EP3248541A1) and CE-marked chest-worn sensor (Biocubica srl, Milan, Italy), named Soundi$^\text{\textregistered}$. It allows for the simultaneous recording of electrocardiographic,  photoplethysmographic, acoustic, accelerometer, bioimpedance, and temperature (body surface and ambient) signals for a duration of about 24~hrs. It features a circular shape of diameter not as much as 6 cm and a thickness of approximately 1 cm, with a weight not exceeding 40 grams. Medically certified double-sided tape ensures that it sticks securely to the chest. Further technical details and performance can be found in \cite{Marzorati2020, Rossi2022}. 

\subsection{Acquisition protocol}
Two identical Soundi systems were adopted, one to collect the PPG signal on the chest (cPPG), considered to feature low quality and undergoing restoration, and another sensor to obtain a high-quality PPG signal measured on the finger (fPPG), considered the reference signal. The first sensor was positioned on the chest surface close to the heart, on the 2nd-3rd intercostal space. A medically certified, double-sided adhesive tape securely affixes the device to the chest, enabling the acquisition of the 3-channel cPPG signal. Simultaneously, the subject was asked to put his/her left ring finger in contact with the optical sensor of the second device, acquiring the 3-channel fPPG signal. The adherence between the finger and the optical sensor was verified to avoid any movement or environmental light artifact effects. Additionally, the finger was covered to minimize interference from external light.
Fourty healthy adult subjects (20 males and 20 females), aged 27.02$\pm$9.34 years (27.40$\pm$8.97 for males and 26.67$\pm$10.11 for females), with no prior cardiac conditions, were enrolled in the study. 
The acquisition protocol included a 10-minute rest period in sitting posture. For each participant, on average 3 repetitions of the acquisition protocol were performed. Between each participant, proper sanitization of the utilized devices was carried out. Data was collected under the Ethical Committee approval Opinion 3/2019, Politecnico di Milano University. For the first 30 subjects, only the PPG signal was recorded. For testing purposes, the last ten subjects underwent the concurrent recording of a single-lead (lead II) ECG, using the Soundi system as well. The onboard-acquired fPPG and cPPG signals were downloaded to a PC and uniformly sampled at 400 Hz. Both signals were processed using a 4th-order band-pass Bessel filter with a frequency range of 0.5 to 25 Hz. The Bessel filter was chosen over other digital filters due to its superior performance in minimizing both waveform distortion and phase shift \cite{Lapitan2024}. Each recording was segmented into 5-second windows (referred to as chunks), containing 2000 samples each. To increase the number of chunks, a 2-second overlap was applied between consecutive windows. Importantly, each 5-second fPPG chunk was matched with its corresponding 5-second cPPG chunk. Thanks to the overlap, each recording had 150 chunk pairs. In total, 16500 chunks pairs were attained. 

\begin{figure}[t!]
    \centering
    \includegraphics[scale=0.6]{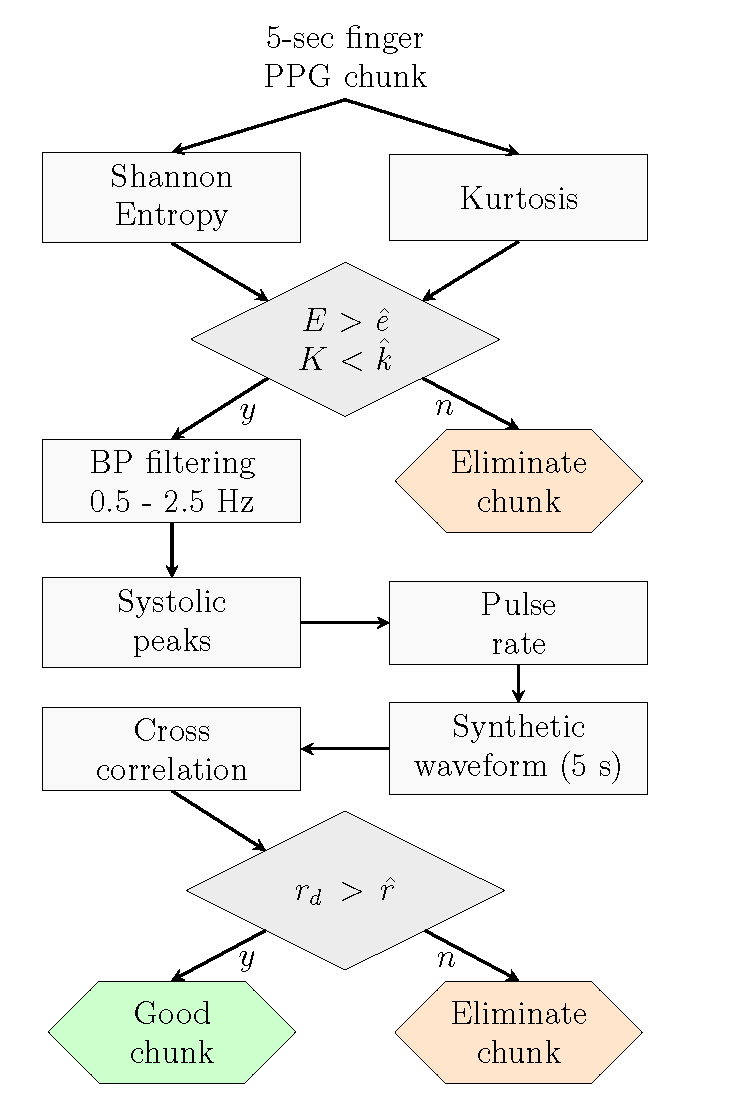}
    \caption{Procedure to label fPPG 5-s chunks.}
    \label{fig:pre_proc}
\end{figure}

\begin{figure}[t!]
    \centering
    \includegraphics[width=1\linewidth]{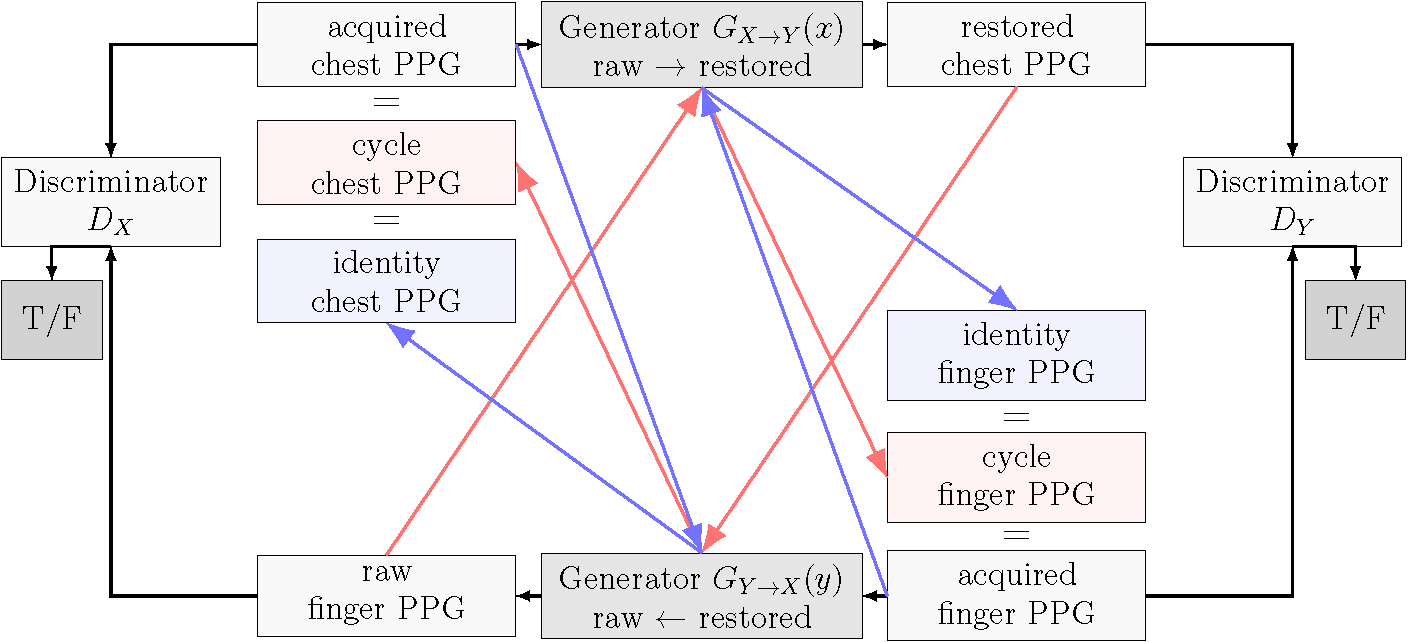}
    \caption{Architecture of the starGAN model.}
    \label{fig:netschema}
\end{figure}

\subsection{Quality check and chunk labeling}
The quality check of the 5-second chunks was developed by verifying first the fPPG and later the cPPG. One fPPG chunk was retained only whether all the three fPPG waveforms were considered of acceptable quality, irrespective of the corresponding cPPG quality. The method was developed using a template matching procedure (Fig. \ref{fig:pre_proc}). 
Shannon $E$ entropy (calculated with 16 bins) and Kurtosis $K$ indices were first computed for each fPPG chunk and a threshold check was then applied, based on the concurrent conditions $K$$<$3.5 and $E$$>$0.8 \cite{Selvaraj2011}. Signal chunks meeting these thresholds were subjected to aggressive filtering using a 4th-order Butterworth filter with a passband of 0.5–2.5 Hz to perform peak extraction~\cite{Elgendi2013}, followed by pulse rate calculation (Eq. \ref{eq:PR}), independently for all three finger channels. Using the derived pulse rates, which may vary across channels in noisy conditions, a PPG reference signal was generated by replicating a nominal pulse waveform into the 5-second time window to reflect the specific computed pulse rate on the chunk under analysis. This was accommodated for all the three channels.
To evaluate both the alignment and similarity between the target fPPG and the corresponding template signals, we computed the cross-correlation after rescaling both signals to the range [0, 1]. The normalized correlation value, $r_d$, at the optimal time delay $d$, was given by: 

\begin{equation}
     r_d = \frac{\sum_i({(t[i]-\hat{t})(s[i-d]-\hat{s})})}{\sqrt{\sum_i(t[i]-\hat{t})^2} \sqrt{\sum_i(s[i-d]-\hat{s})^2}}
\end{equation}

\noindent where $\hat{t}$ and $\hat{s}$ were the average value of the template waveform and the average value of the signal waveform, respectively. A threshold $\hat{r}$ equal to 0.8 for $r_d$ was selected to separate acceptable from unacceptable chunks (cfr. Fig. \ref{fig:pre_proc}). Once the fPPG chunk was properly labeled, the corresponding cPPG was reviewed using a visualization dashboard. Each channel within the chunk was manually labeled as either 'keep' or 'leave'. A channel was labeled as 'keep' if a visible periodicity allowed the identification of the number of beats within the chunk. Conversely, a channel was labeled as 'leave' if no clear periodicity could be observed.
A chunk was retained only if at least one of its channels was labeled as 'keep'. The rationale behind this selection process is that the network requires the presence of discernible periodicity in at least one channel to reconstruct a clean signal. If none of the channels exhibit periodic beats, the network would lack a clear reference for what a clean version of the signal should look like, rendering the reconstruction process unreliable.
Overall, about 5113 valid chunk pairs were retained. All the 5113 valid chunks obtained have been subdivided into three subsets: train set, validation set and test set, in a subject-wise manner. Respectively, the train was made of the valid chunks relative to the first 20 subjects, the validation set had the valid data coming from 10 other subjects and the test set was composed of the valid chunks relative to the last 10 subjects. For training purposes, data augmentation was performed by generating random Gaussian noise with a zero mean and a standard deviation of 50, based on a careful inspection of the noisy generated chunks. This noise was filtered within the same frequency range as the processed PPG signal and then added to each measured cPPG chunk in the training set. As a result, the total number of chunks in the training set was increased to 6026. The total number of chunks for the validation set and the test set was equal to, respectively, 1008 and 1092. Therefore, the total number of chunks after the data augmentation was equal to 8126.

\subsection{starGAN Network}
The developed network, named starGAN, was based on a cycle coherent GAN architecture. It comprised two main components, namely the generator, responsible for signal-to-signal mapping, and the discriminator enabling the signal classification. Two generator-discriminator couples were defined into the cyclic architecture, one to map noised to restored signals and the other to map back restored to noised signals (Fig. \ref{fig:netschema}). The interaction between these elements worked as follows. The generator $G_{X \to Y}(x)$ processed the acquired chest PPG in the domain $X$ and produced a corresponding restored chest PPG in the domain $Y$. The discriminator $Y$ alternatively evaluated the restored chest signal and the corresponding acquired finger PPG signal, trying to label original and restored signals as ``1'' and ``0'', respectively. During the training, the goal of generator $G_{X \to Y}(x)$ was to progressively increase the quality of the restored signal to deceive the discriminator $Y$ into mistaking reconstructed chest PPG for real finger PPG. Likewise, generator $G_{Y \to X}(y)$ took the acquired finger PPG and generated a noised version of it. Discriminator $X$ distinguished between acquired chest PPG (``1'') and noised finger PPG (``0''). This basic structure was extended with two additional cycles to ensure cyclic coherence. To reinforce the style-transfer process, the chest PPG restored by generator $G_{X \to Y}(x)$, was fed into generator $G_{Y \to X}(y)$. The corresponding output was the  ``cycle'' chest PPG expected to match the acquired chest PPG. An identical process occurred for the noised version of the finger PPG, being converted back into a “cycle” denoised finger PPG to be compared with the original acquired finger PPG. The two generators were both implemented through an identical UNet model, featuring encoder and decoder branches along with skip connections. Model training was executed on a virtual machine with 64 GB of RAM and 64 CPU cores, without the use of GPUs. The implementation was carried out using Python v.3.10 and TensorFlow v.2.14.

\subsubsection{Generator}
The baseline generator architecture consisted of an encoder-decoder structure with three convolutional blocks in the encoder. Each block included a 1D convolution, instance normalization, and a Swish activation function. The encoder processed the three input channels (red, green, and infrared), with the first block producing 16 feature maps. The number of feature maps doubled at each subsequent block, reaching 64 at the bottleneck. Downsampling was performed between consecutive blocks using additional convolutional layers with a stride of 2. The bottleneck featured an extra convolutional layer that maintained the same number of input and output feature maps while preserving the signal length. The kernel size for all convolutions was fixed at 9. The decoder mirrored the encoder, using 1D transposed convolutions between consecutive blocks to restore the original signal length. Skip connections were incorporated between corresponding encoder and decoder layers, following the standard UNet architecture. A final convolutional layer with a linear activation function produced the three  output channels. The input to each generator was a three-dimensional array of 2000 samples, corresponding to a 5-second signal recorded at 400 Hz. The baseline discriminator architecture consisted of an encoder with five convolutional blocks (which had the same structure as the convolutional blocks in the generators) and a final convolutional layer, all with a kernel size of 7. Variations of the baseline generator architecture were explored by modifying the number of feature maps, the number of convolutional blocks, and their respective kernel sizes, including configurations with variable kernel sizes. Additional modifications included the integration of inception convolutional blocks. Each inception block consisted of three parallel convolutional paths, with their resulting feature maps concatenated to form the block's output. Gated recurrent unit (GRU) layers within the skip connections, and bidirectional long short-term memory (BiLSTM) layers in the network output were also investigated. 

\subsubsection{Discriminator}
The discriminator processed the three-channel input signal and utilized a sigmoid activation function in its final layer for classification. Starting with 8 feature maps in the first block, the number of feature maps increased progressively, reaching 64 in the fourth convolutional block. In the fifth block, the number of feature maps was halved to 32, and the final convolutional layer reduced this further to a single output, which was passed through the sigmoid activation function for classification. For the discriminator, variations focused on altering the number of feature maps and kernel sizes. In total, 11 distinct architectures were designed, implemented, and trained (Table \ref{tab:Models}). 

\begin{table}[t!]
    \centering
    \caption{Eleven different architectures. G I. stands for the initial number of filters of the generator network, D I. indicates the initial number of filters of the discriminator network. Similarly G K. and D K. represent the kernel sizes of the generator and the discriminator, respectively. (y: yes; n: no; fixed: G K. = 9, D K. = 7; var: G K. = 9-5-5, D K. = 9-7-5-3-1)}
    \label{tab:Models}
    \begin{tabular}{c c c c c c c c}
        \toprule
        & G I. & D I. & Inception & G K. & D K.    & GRU & LSTM\\
        \midrule
        m01 & 8 & 4 & y(1st) & fixed & fixed & n & n \\
        m02 & 16 & 8 & y(1st) & fixed & fixed & n & n \\
        m03 & 32 & 16 & y(1st) & fixed & fixed & n & n \\
        m04 & 16 & 8 & n & fixed & fixed & n & n \\
        m05 & 16 & 8 & y(all) & fixed & fixed & n & n \\
        m06 & 16 & 8 & y(1st) & var & var & n & n \\
        m07 & 16 & 8 & y(1st) & var & fixed & n & n \\
        m08 & 16 & 8 & y(1st) & fixed & var & n & n \\
        m09 & 16 & 8 & y(1st) & fixed & fixed & y & n \\
        m10 & 16 & 8 & y(1st) & fixed & fixed & n & y \\
        m11 & 16 & 8 & y(1st) & fixed & fixed & y & y \\
        \bottomrule
    \end{tabular}
\end{table}

\subsection{Loss functions and training}
For the training, the weights of the starGAN were computed by minimizing a multi-domain loss, one for each branch. Along with the adversarial and cycle losses, an identity loss was considered as well to preserve physiological features of the PPG domain, such as systolic peak amplitude, cardiac rate and signal variability, and avoiding spurious components introduced by the generator.
The adversarial loss ensured that the generators produce signals indistinguishable from the real ones. The adversarial loss $\mathcal{L}_{\text{adv}}^Y$, corresponding to the restored cPPG domain $Y$, was composed by two terms as: 

\begin{equation}
\mathcal{L}_{\text{adv}}^{Y,G}(G_{X \to Y}, D_Y) = \mathbb{MSE}[1, D_Y(G_{X \to Y}(x))]
\end{equation}

\noindent to estimate the weights of the generator $G_{X \to Y}$, where $\mathbb{MSE}$ is the mean squared error, and

\begin{equation}
\mathcal{L}_{\text{adv}}^{Y,D}(G_{X \to Y}, D_Y) = \mathbb{MSE}[0, D_Y(G_{X \to Y}(x))] + \mathbb{MSE}[1, D_Y(y)]
\end{equation}

\noindent to estimate the weights of the discriminator $D_Y$. The adversarial loss $\mathcal{L}_{\text{adv}}^X$, corresponding to the measurement cPPG domain $X$, was composed by two terms as:

\begin{equation}
\mathcal{L}_{\text{adv}}^{X,G}(G_{Y \to X}, D_X) = \mathbb{MSE}[1, D_X(G_{Y \to X}(y))]
\end{equation}

\noindent to estimate the weights of the generator $G_{Y \to X}$ and

\begin{equation}
\mathcal{L}_{\text{adv}}^{X,D}(G_{Y \to X}, D_X) = \mathbb{MSE}[0, D_X(G_{Y \to X}(y))] + \mathbb{MSE}[1, D_X(x)]
\end{equation}

\noindent to estimate the weights of the discriminator $D_X$. The cyclic consistency loss ensured the signal restoration after a forward and backward transformation. For the two cycles $X \to Y \to X$, and $Y \to X \to Y$, the losses were:

\begin{equation}
\mathcal{L}_{\text{cycle}}^X(G_{X \to Y}, G_{Y \to X}) = \mathbb{E}_{x \sim p_X}[\| G_{Y \to X}(G_{X \to Y}(x)) - x \|] 
\end{equation}

\begin{equation}
\mathcal{L}_{\text{cycle}}^Y(G_{Y \to X}, G_{X \to Y}) = \mathbb{E}_{y \sim p_Y}[\| G_{X \to Y}(G_{Y \to X}(y)) - y \|].
\end{equation}

\noindent where $\|\cdot\|$ is the absolute value. The identity losses for the generators $G_{X \to Y}$ and $G_{Y \to X}$, the two losses were:

\begin{equation}
\mathcal{L}_{\text{id}}^Y(G_{X \to Y}) = \mathbb{E}_{y \sim p_Y}[\| G_{X \to Y}(y) - y \|] 
\end{equation}

\begin{equation}
\mathcal{L}_{\text{id}}^X(G_{Y \to X}) = \mathbb{E}_{x \sim p_X}[\| G_{Y \to X}(x) - x \|]
\end{equation}

\noindent The total losses for the two generators combine the three components as:
\begin{equation}
\mathcal{L}_{\text{total}}^{Y}(G_{X \to Y}, D_Y) = \lambda_{\text{adv}} \mathcal{L}_{\text{adv}}^{Y,G} + \lambda_{\text{cycle}} \mathcal{L}_{\text{cycle}}^Y + \lambda_{\text{id}} \mathcal{L}_{\text{id}}^Y,
\end{equation}
\begin{equation}
\mathcal{L}_{\text{total}}^{X}(G_{Y \to X}, D_X) = \lambda_{\text{adv}} \mathcal{L}_{\text{adv}}^{X,G} + \lambda_{\text{cycle}} \mathcal{L}_{\text{cycle}}^X + \lambda_{\text{id}} \mathcal{L}_{\text{id}}^X,
\end{equation}
where $\lambda_{\text{adv}}, \lambda_{\text{cycle}}, \lambda_{\text{id}}$ were hyperparameters controlling the relative importance of each term. According to the literature, the three terms were set to 1, 10, 5, respectively \cite{Zhu2017}.

\subsection{Signal quality metrics}
\label{sec: metrics}
To evaluate the performance of different models, various metrics were assessed, encompassing time and frequency domains. Calculating the time-domain metrics required aligning the measured and restored cPPG signal with the fPPG signal due to their phase differences which arises from the distinct positions of the sensors. To address this, the cross-correlation was computed between the restored cPPG and fPPG signals, and the lag that maximized this cross-correlation was used to shift both the measured and restored cPPG signals relative to the fPPG signal. This approach ensured a uniform shift for both the measured and restored cPPG signals relative to the fPPG. Any non-overlapping portions of the signal were subsequently removed. The time-domain metrics included root mean squared error ($RMSE_t$) in normalized units, mean absolute error ($MAE$) in normalized units, Pearson’s correlation coefficient ($R$), and the coefficient of determination ($R^2$). The signal-to-noise ratios ($SNR$) for both the measured and restored signals were computed as:

\begin{align}
    SNR_{m} = 10log_{10}\left( \frac{\sigma^{2}_{fPPG}}{\sigma^{2}_{m}} \right) \\
    SNR_{r} = 10log_{10}\left( \frac{\sigma^{2}_{fPPG}}{\sigma^{2}_{r}} \right) 
\end{align}

\noindent where $\sigma^{2}_{m}$ and $\sigma^{2}_{r}$ represent the noise power for the measured and restored signals, respectively. These were computed as the variance of the differences cPPG-fPPG for the measured case and restored cPPG-fPPG for the restored case. 

In the frequency domain, the root mean squared error between the target signal $y_i$ and the measured/restored signal $k_i$ was calculated according to \cite{Mahmud2024} as:

\begin{equation}
     RMSE_f = \sqrt{\frac{1}{N}\sum{(PSD(y_i) - PSD(k_i))^2}}
\end{equation}
\noindent where \textit{PSD} indicates the power spectral density, which was computed using Welch method using a Hann's window equal to 10\% of the chunk length (200 samples).

\subsection{Clinical Metrics}
\subsubsection{Restored cPPG agaists fPPG}
The peak to peak ($PP$) temporal intervals and the pulse rate ($PR$) from the PPG were calculated as primary clinical metrics, following the alignment of the restored and measured cPPG with the fPPG as described above. Systolic peaks in each chunk were automatically computed \cite{Bishop2018}.
Given the systolic peaks of a chunk, the $PP_{chunk}$ interval, computed in seconds, along with the corresponding pulse rate $PR_{chunk}$ (bpm), were calculated as the average of consecutive intervals $\delta{p_i} = (p_{i+1} - p_i)$ within the considered chunk:

\begin{align}
    PP_{chunk} = \frac{t_s}{N_{p}} \sum_{i=0}^{N_{p}-1} \delta{p_i},   PR_{chunk} = \frac{60}{PP_{chunk}}
    \label{eq:PR}
\end{align}

\noindent where $t_s$ and $N_{p}$ represent the sampling interval and the number of systolic peaks in the chunk, respectively. For the $PR$, the mean absolute error $MAE_{PR}$ was computed. 
Linear regression analysis was performed for the two indices by comparing the fPPG with both the original and the predicted cPPG. The Pearson correlation coefficients $R_{PP}$ and $R_{PR}$ were computed to evaluate the agreement between the measured cPPG and the reference fPPG, as well as between the restored cPPG and reference fPPG.
\subsubsection{Restored cPPG against ECG}
For the 10 subjects in the test set, additional analyses were conducted using the available ECG data. The correlation coefficient $R_{RR}$  was computed between $PP_{chunk}$, derived from both the measured and restored cPPG signals, and $RR_{chunk}$, derived from the ECG. Similarly, the correlation coefficient $R_{HR}$ was calculated between $PR_{chunk}$, derived from both the measured and restored cPPG signals, and $HR_{chunk}$, derived from the ECG signal.
The R peaks, relative to the $ECG_{chunk}$ signal, were automatically computed \cite{Nabian2018}. The $RR$ intervals and $HR$ values were calculated similarly to eqs. (\ref{eq:PR}). For $HR$, the mean absolute error $MAE_{HR}$ was computed as well. 

\section{Experimental sessions and results}
\subsection{Hyperparameter ablation}
\begin{figure*}[t!]
    \centering
    \includegraphics[width=1\linewidth]{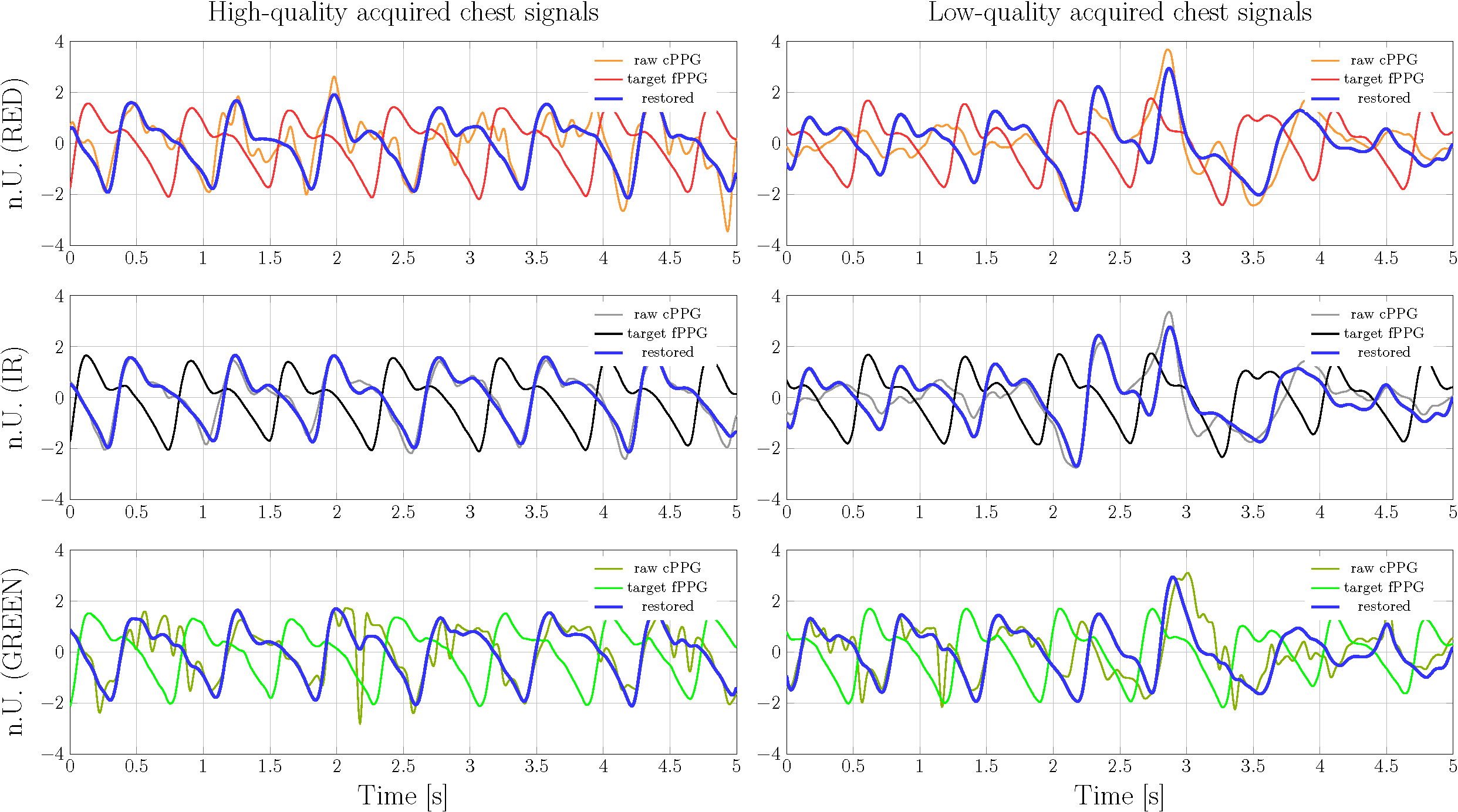}
    \caption{Restoration of high- (left panel) and low-quality (right panel) measured cPPG signals. In both cases the pulse rate can be easily detected from the restored cPPG signal (blue line). More specifically, in the restoration of high-quality signals, the R values between the restored cPPG signal and the fPPG signal reached values of 0.98, 0.99, and 0.99 for the red, infrared, and green channels, respectively. In contrast, for the low-quality signal restoration, the R values were 0.63, 0.62, and 0.70 for the red, infrared, and green channels, respectively.} 
    \label{fig:H_Q_signals}
\end{figure*}

The evaluation of the 11 different architectures  revealed varying levels of performance (Table \ref{tab:ablation_results}). A statistical analysis of the three (R$_t$, IR$_t$, and G$_t$) temporal $RMSE$ indices was conducted using the Kruskal-Wallis non-parametric test, followed by a post-hoc comparison with Bonferroni adjustment (significance level: $p$ = 0.01). Architecture m09 achieved the lowest temporal $RMSE_t$ average value (0.47) and demonstrated statistically significant differences compared to all other architectures, except for m02 ($p$=0.31) and m05 ($p$=0.18). The analysis of frequency $RMSE$ average values confirmed the lowest value (0.0022) for the architecture m09, with no statistical difference from m02 ($p$=0.20).  Both Pearson correlation $R$ and coefficient of determination $R^2$ were coherent with $RMSE$ values, achieving maximum values with m09, namely 0.89 and 0.78, respectively. Accordingly, architecture m09 was selected as optimal for the next evaluation on the test set. 
As an example, two chunks were depicted to illustrate the model's ability to restore chest PPG with different original qualities of the measured signal (Fig. \ref{fig:H_Q_signals}). In the left panel (high-quality chest PPG), the chest IR signal exhibited both clear beat periodicity and optimal waveform quality, resulting in effective restoration across all three channels in terms of both periodicity and waveform shape. In contrast, the right panel (low-quality chest PPG) shows that the green measured cPPG presents beat periodicity despite significant distortion in the waveform. The model successfully restored periodicity across all three channels and improved the waveform shape, although it remained suboptimal. Despite this, the pulse rate could still be reliably detected.

\begin{table*}[t]
    \centering
    \caption{Ablation results (median values). R$_t$, IR$_t$ and G$_t$ are the $RMSE$ indices in the time domain, respectively, for red, IR, and green channels. AV$_t$ is the corresponding average value. R$_f$, IR$_f$ and G$_f$ are the $RMSE$ indices in the frequency domain, respectively, for red, IR, and green channels. AV$_f$ is the corresponding average value. $R$ and $R_2$ values are reported as well.} 
    \label{tab:ablation_results}
     \small
     \begin{tabular}{c c c c c c c c c c c c c c c c c}
        \toprule
        & R$_t$ & IR$_t$ & G$_t$ & AV$_t$ & R$_f$ & IR$_f$ & G$_f$ & AV$_f$ & R$_R$ & IR$_R$ & G$_R$ & AV$_R$ & R$_R2$ & IR$_R2$ & G$_R2$ & AV$_R2$\\        
        \midrule
        m01 & 0.52 & 0.51 & 0.51 & 0.51 & 2.3e-3 & 2.5e-3 & 2.4e-3 & 2.4e-3 & 0.87 & 0.87 & 0.87 & 0.87 & 0.73 & 0.74 & 0.74 & 0.74 \\
        m02 & 0.50 & 0.49 & 0.46 & 0.48 & 2.4e-3 & 2.4e-3 & 2.2e-3 & 2.3e-3 & 0.88 & 0.88 & 0.90 & 0.89 & 0.75 & 0.76 & 0.79 & 0.77 \\
        m03 & 0.54 & 0.51 & 0.48 & 0.51 & 2.5e-3 & 2.5e-3 & 2.4e-3 & 2.5e-3 & 0.86 & 0.87 & 0.89 & 0.87 & 0.71 & 0.74 & 0.77 & 0.74 \\
        m04 & 0.56 & 0.55 & 0.51 & 0.54 & 2.6e-3 & 2.8e-3 & 2.8e-3 & 2.7e-3 & 0.84 & 0.85 & 0.87 & 0.85 & 0.68 & 0.70 & 0.74 & 0.71 \\
        m05 & 0.51 & 0.49 & 0.47 & 0.49 & 2.4e-3 & 2.5e-3 & 2.3e-3 & 2.4e-3 & 0.87 & 0.88 & 0.89 & 0.88 & 0.74 & 0.76 & 0.78 & 0.76 \\
        m06 & 0.62 & 0.59 & 0.56 & 0.59 & 3.7e-3 & 3.2e-3 & 3.2e-3 & 3.4e-3 & 0.81 & 0.83 & 0.85 & 0.83 & 0.63 & 0.65 & 0.69 & 0.66 \\
        m07 & 0.67 & 0.62 & 0.57 & 0.62 & 3.6e-3 & 3.4e-3 & 3.8e-3 & 3.6e-3 & 0.78 & 0.81 & 0.84 & 0.81 & 0.56 & 0.61 & 0.68 & 0.62 \\
        m08 & 0.51 & 0.49 & 0.49 & 0.49 & 2.4e-3 & 2.5e-3 & 2.6e-3 & 2.5e-3 & 0.88 & 0.88 & 0.88 & 0.88 & 0.75 & 0.76 & 0.77 & 0.76 \\
        m09 & 0.49 & 0.47 & 0.45 & 0.47 & 2.3e-3 & 2.2e-3 & 2.1e-3 & 2.2e-3 & 0.88 & 0.89 & 0.90 & 0.89 & 0.76 & 0.79 & 0.80 & 0.78 \\
        m10 & 0.53 & 0.51 & 0.49 & 0.51 & 6.8e-3 & 7.4e-3 & 6.7e-3 & 7.0e-3 & 0.85 & 0.87 & 0.88 & 0.86 & 0.72 & 0.74 & 0.76 & 0.74 \\
        m11 & 0.55 & 0.54 & 0.54 & 0.54 & 8.9e-3 & 9.0e-3 & 9.0e-3 & 9.0e-3 & 0.84 & 0.85 & 0.84 & 0.85 & 0.71 & 0.72 & 0.71 & 0.71 \\
        \bottomrule
    \end{tabular}
\end{table*}

\subsection{Validation on the test set}
\begin{figure*}[t!]
    \centering
    \includegraphics[width=0.85\linewidth]{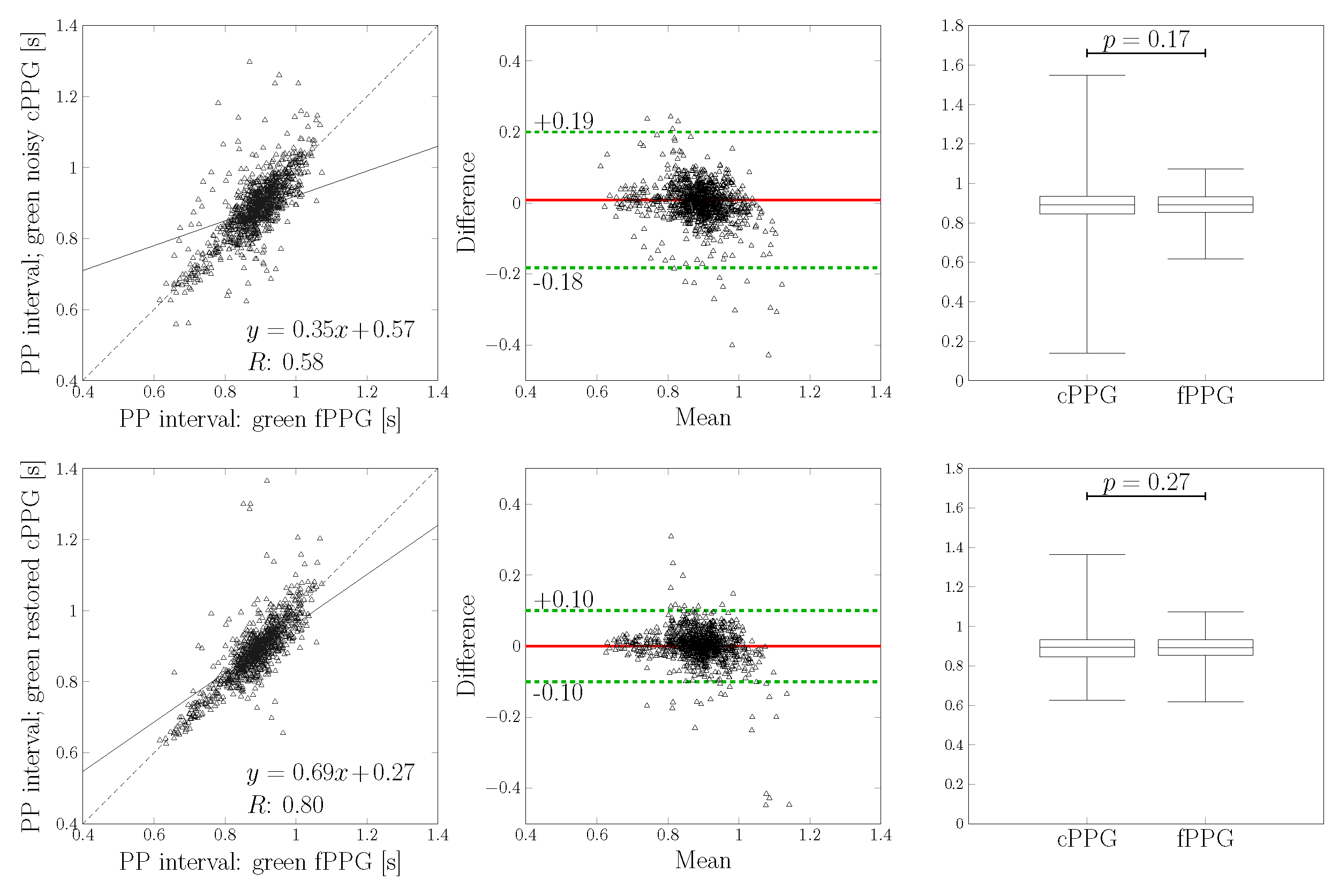}
    \caption{Scatterplot and Bland-Altman for test set. Without lack of generality the green channel was here considered. Top: plots for the PP intervals (s) of measured cPPG vs. fPPG, along with the corresponding PP distributions and the $p$ value. Bottom: plots for the PP intervals of restored cPPG vs. fPPG, along with the corresponding PP distributions and the $p$ value.}
    \label{fig:blandaltman}
\end{figure*}

Once model m09 was identified as the best-performing architecture, it was retrained using the combined training and validation datasets. The final optimized model was subsequently evaluated on the test set. This model comprised a total of 140,567 parameters and required approximately 24 hours to complete training on a CPU-only system. The inference time for each chunk was approximately 0.07 seconds. In the frequency domain, $RMSE$ values (Table \ref{tab:signal_quality_test}) demonstrated an average decrease of 50\% across the three channels. In the time domain, the metrics showed an average reduction of about 38 and 37\% for $MAE$ and $RMSE$, respectively. $SNR$, $R$ and $R^{2}$ indices improved on average of about 125, 24 and 80\%, respectively.
The clinical indices all revealed relevant improvements after signal restoration (Table \ref{tab:clinical_quality_test}).  $R_{PR}$ and $R_{HR}$ values increased by three times, while $R_{PP}$ nearly doubled its value. Likewise, the $MAE_{PR}$  (restored vs measured) decreased on average along all the three channels from 7 to 2~bpm. Focusing on the peak to peak interval, the restoration was effective  according the Bland-Altman plot (Fig. \ref{fig:blandaltman}). Without lack of generality, results for the green channel (cfr. Table \ref{tab:clinical_quality_test})
showed an improvement of the difference from about 0.37 to 0.20~s (1.96~SD). While no statistically significant differences among the 2 $PP$ distributions were found, the variability increased of about 42\%. The comparison of the pulse rate, computed using the restored PPG, against the pulse rate computed with the acquired ECG provided a $MAE_{HR}$ of about 2~bpm. 

\begin{table}[t]
    \centering
    \caption{Median values of signal quality metrics on the test set (meas = measured, res = restored).}
    \label{tab:signal_quality_test}
    \begin{tabular}{c | c c c c c c}
        \toprule
        &
        \multicolumn{3}{c}{\textbf{meas cPPG vs fPPG}}  &  \multicolumn{3}{c}{\textbf{res cPPG vs fPPG}} \\ 
         \midrule
          Metrics & RED & IR & GREEN & RED & IR & GREEN \\
        \midrule
        $RMSE_f$ & 5e-3 & 3.6e-3 & 5.5e-3 & 2.4e-3 & 2.3e-3 & 2.4e-3\\
        $RMSE_t$ & 0.87 & 0.66 & 0.72 & 0.47& 0.46 & 0.47 \\
        $MAE$ & 0.68 & 0.51 & 0.56 & 0.36 & 0.36 & 0.37 \\
        $R$ & 0.63 & 0.78 & 0.75 & 0.89 & 0.90 & 0.89 \\
        $R^2$ & 0.25 & 0.57 & 0.49 & 0.78 & 0.79 & 0.78 \\
        $SNR$ & 2.10 & 3.46 & 3.08 & 6.28 & 6.50 & 6.43 \\
        \bottomrule
    \end{tabular}
\end{table}

\begin{table}[t]
    \centering
    \caption{Clinical quality indices on the test set. The reported values are median values (meas = measured, res = restored). $MAE_{PR}$ index is expressed in bpm.}
    \label{tab:clinical_quality_test}
    \begin{tabular}{c | c c c c c c}
        \toprule
        &
        \multicolumn{3}{c}{\textbf{meas cPPG vs fPPG/ECG}}  &  \multicolumn{3}{c}{\textbf{res cPPG vs fPPG/ECG}} \\ 
         \midrule
          Metrics & RED & IR & GREEN & RED & IR & GREEN \\
        \midrule
        $R_{PP}$  & 0.17 & 0.44 & 0.58 & 0.72 & 0.78 & 0.80 \\
        $R_{PR}$  & 0.06 & 0.50 & 0.26 & 0.80 & 0.84 & 0.85 \\
        $R_{HR}$  & 0.07 & 0.53 & 0.26 & 0.81 & 0.85 & 0.87 \\
        $R_{RR}$  & 0.19 & 0.48 & 0.61 & 0.73 & 0.80 & 0.83 \\
        $MAE_{PR}$ & 11 & 4 & 6 & 2 & 2 & 2 \\
        \bottomrule
    \end{tabular}
\end{table}

\subsection{Effects on the time shift in the restored cPPG}
\begin{figure}[t!]
    \centering
    \includegraphics[width=0.7\linewidth]{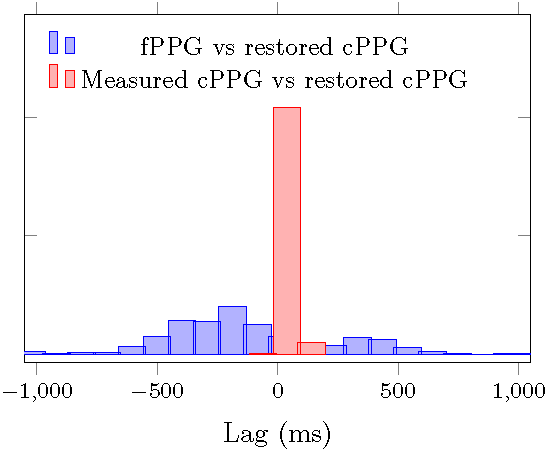}
    \caption{Lag histogram for the restored cPPG against measured cPPG and fPPG.}
    \label{fig:lags}
\end{figure}

In order to assess whether the starGAN affected the timing of the restored cPPG introducing possible undue shifts with respect to the acquired PPG at the chest, we measured the lags that maximized the cross-correlation with both the original measured cPPG and the fPPG (Fig. \ref{fig:lags}). As expected, the restored cPPG kept a high temporal agreement with the acquired cPPG, providing an average time shift of about -0.6~ms (SD=23.7~ms). For comparison, the restored chest PPG against the finger PPG featured an average time shift of about -64~ms (SD=439~ms). 

\subsection{Sensor fusion effectiveness}
\begin{figure*}[h]
    \centering
    \includegraphics[width=\linewidth]{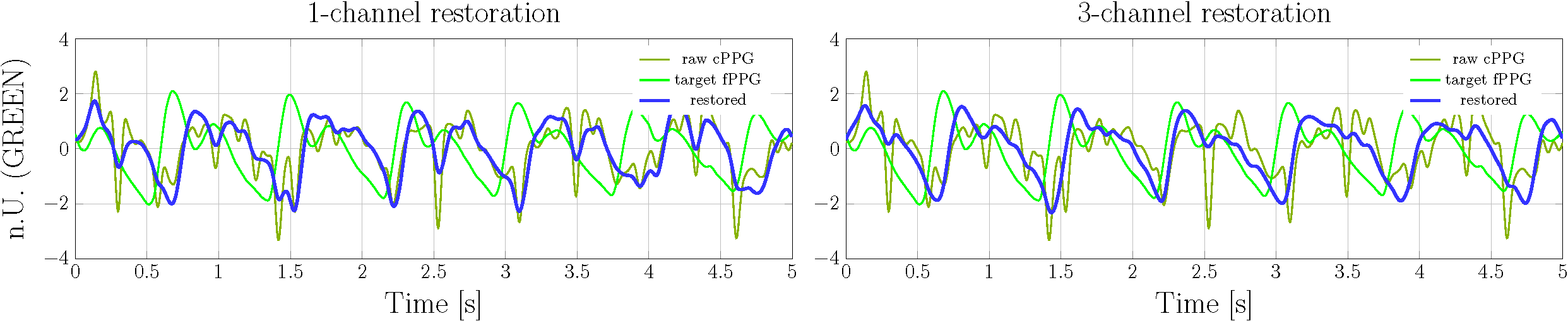}
    \caption{Restoration of single-channel model against three-channel model.}
    \label{fig:1vs3channels}
\end{figure*}

The objective of this experiment was to assess the advantages of utilizing all three channels compared to relying solely on one channel. Specifically, the m09 architecture was adapted to handle single 1-D input and output data and subsequently trained. Without lack of generality, the analysis was performed on the green channel. Both signal quality metrics  (Table \ref{tab:signal_quality_3channels_vs_1channel}) and clinical indices (Table \ref{tab:clinical_quality_3channels_vs_1channel}) demonstrated inferior performance for the single-channel model. Remarkably, time and frequency $RMSE$ increased of about 29\% and 17\% respectively. Similarly, the $MAE$ increased by about 11\%. The $SNR$ decreased from 6.43 to 5.85 (about 9\%).
Likewise, clinical quality metrics such as $R_{PR}$ and $R_{HR}$ each deteriorated by roughly 9\%. The $MAE_{PR}$ decreased from 2 to 3 bpm.
The graphical comparison on the same processed chunk showed clear evidence of the poorer quality of the green PPG signal restored using the single-channel model (Fig. \ref{fig:1vs3channels}).

\begin{table}[t!]
    \centering
    \caption{Signal quality metrics (median values) on the test set for the GREEN channel: comparison between reconstructions using three input channels vs. a single input channel (meas = measured, res = restored.}
    \label{tab:signal_quality_3channels_vs_1channel}
    \begin{tabular}{c | c c c}
        \toprule
        & \textbf{meas cPPG vs fPPG} &  \multicolumn{2}{c}{\textbf{res cPPG vs fPPG}} \\ 
         \midrule
          Metrics &  & 1 channel &  3 channels \\
        \midrule
        $RMSE_f$ & 5.5e-3 & 3.1e-3  & 2.4e-3\\
        $RMSE_t$  & 0.72 & 0.55  & 0.47 \\
        $MAE$  & 0.56 & 0.42  &  0.37\\
        $R$ & 0.75  & 0.85  & 0.89\\
        $R^2$  & 0.49 & 0.69  & 0.78 \\
        $SNR$ & 3.08 & 5.85  & 6.43\\
        \bottomrule
    \end{tabular}
\end{table}

\begin{table}[t!]
    \centering
    \caption{Clinical quality metrics on the test set for the GREEN channel: comparison between reconstructions using three input channels vs. a single input channel (meas = measured, res = restored). }
    \label{tab:clinical_quality_3channels_vs_1channel}
    \begin{tabular}{c c c c}
        \toprule
        &
        \textbf{meas cPPG vs fPPG/ECG}  &  \multicolumn{2}{c}{\textbf{res cPPG vs fPPG/ECG}} \\ 
         \midrule
          Metrics & GREEN & G\_3CH & G\_1CH \\
        \midrule
        $R_{PP}$   & 0.59 & 0.80 & 0.76 \\
        $R_{PR}$   & 0.26 & 0.85 & 0.77 \\
        $R_{HR}$   & 0.26 & 0.87 & 0.79 \\
        $R_{RR}$   & 0.61 & 0.83 & 0.78 \\
        $MAE_{PR}$ & 6   &  2     & 3 \\
        \bottomrule
    \end{tabular}
\end{table}

\section{Discussion} 
The study focused on restoring chest PPG signals, collected by wearable devices, addressing their inherent weakness and noise compared to finger PPG \cite{Park2022}. We showcased that the proposed starGAN effectively restored the chest PPG waveforms, sensibly increasing the correlation with the concurrent reference finger PPG signal, while preserving the original timing. The model significantly enhanced heart rate estimation when compared to ECG-derived heart rate. The best-performing architecture (m09) reduced the signal $RMSE$ by 37\% in the time domain and 50\% in the frequency domain while tripling the accuracy of clinical metrics such as the pulse rate correlation. We also confirmed the advantages of multi-channel PPG sensor fusion, as the restoration process using three input channels outperformed single-channel method (cfr. Table \ref{tab:clinical_quality_3channels_vs_1channel}).  The dual-sensor testing protocol showcased significant advantages over relying on simulated noise for assessing improvements provided by the starGAN. Unlike artificially corrupted signals, which may not accurately capture the complexity and variability of real-world motion artifacts \cite{Lee2019, AfandizadehZargari2023, AvilaCastro2025}, the implemented setup provided authentic noisy and clean reference signals for more effective testing. The approach ensured that the network learns to correct real physiological distortions rather than merely removing artificially injected noise patterns. 
Additionally, it was demonstrated that the starGAN preserved the true temporal and spectral characteristics of PPG signals, preventing over-smoothing and distortion of clinically relevant features such as pulse morphology and heart rate variability. By leveraging synchronized chest and finger PPG signals, the model developed a more robust mapping of motion-induced distortions, leading to improved generalization. 
Compared to state-of-the-art methods (Table \ref{tab:literature}), the proposed starGAN was competitive with most of the previous approaches in restoring the PPG. we not only verify the quality of the clinical related variable but also cared about the PPG waveform. 
In \cite{Lee2019}, the authors trained a BDRAE using MIMIC III dataset, that was artificially noised. The method yielded an $RMSE$ of 6.1 bpm, reporting an improvement of $SNR$ by 7.9~dB. Overall, the noise statistics was trivial and could be effectively handled using a bandpass filter.
Multistage stopband filter was proposed in \cite{Nabavi2020} to reducing motion artifacts reaching a $MAE$ of 0.8~bpm on simulated noisy data. However, the filter parameters required manually setup and the added noises was mainly at frequency higher than physiological signals. 2D-based cycleGAN was proposed to restore the PPG of 33 subjects when corrupted by synthetic noise \cite{AfandizadehZargari2023}, yielding a $RMSE$ of 2.1~bmp. The reported method for corrupting the signals posed questions about the realism of the noise patterns. 
In \cite{Long2024}, the authors, tested a recursive GAN to restore wrist PPG with simulated noise on 15 subjects, achieving a $MAE$ of 1.7~bpm. Again the use of simulated noise, which is hardly realistic limited the generalization of the result to daily life recordings.  
To enhance heart rate estimation, wrist PPG was processed using a 1D-CycleGAN to perform wrist-to-finger PPG translation achieving a correlation between ECG and restored PPG heart rates of about 83\% \cite{Mahmud2024}, which is very similar to our results (cfr Table \ref{tab:clinical_quality_test}). 
Convolutional denoising autoencoders were proposed to restore wrist PPG signals acquired from 48 subjects without requiring additive noise \cite{Mohagheghian2024}. It was reported a reduction in the heart rate root mean square error of about 45\% and an $SNR$ of 4.4. Likewise, GANs with fully connected layers were applied to process MIMIC II waveform database achieving a pulse rate $MAE$ of 1.31~bpm \cite{AvilaCastro2025}. While the simulated noise consisted of sine waves with frequency usually observed in PPG motion artifacts, the normal distribution was unfortunately not realistic to represent the heterogeneity of the artifacts. Despite its challenging performance, the proposed method has some limitations. 
The model was trained and validated on a cohort of healthy young individuals. Its performance on populations with cardiovascular conditions, such as arrhythmias or abnormal perfusion, remains uncertain. Different skin tones and aging effects, impacting optical absorption, were not addressed, which may influence the model's generalizability. In addition, the study primarily involved data collected in a seated posture. Real-world conditions, such as walking, running, or sleeping, introduce more complex motion artifacts that may further degrade performance. Nonetheless, the chest site features lower perfusion, breathing-induced motion, and mechanical interference from thorax movements that may sensibly compromise the signal quality as show in the work.
Further research would be needed to refine the model for various motion conditions, enhance its real-time efficiency, and validate its performance across a broader clinical population. Given that wrist-based PPG signals are similarly prone to quality degradation, the proposed restoration methodology might be tested using smartwatches to enhance the gathered signals.

\begin{table}[t!]
\caption{Comparison of with state-of-the art papers dealing with PPG restoration.}
\centering
\begin{tabular}{lccc}
\midrule
\textbf{Study} & \textbf{Method} & \textbf{Site} & \textbf{Outcome} \\
\midrule
\cite{Lee2019} & BDRAE & wrist &  $SNR$ 7.9, $RMSE$ 6.1 bpm\\
\cite{Nabavi2020} & Multi-stage filter & wrist  & $SNR$ 3.9, $MAE$ 0.8 bpm\\
\cite{AfandizadehZargari2023} & CycleGAN & wrist  & $SNR$ 17.8, $RMSE$ 2.1 bpm\\
\cite{Long2024} & CycleGAN & wrist  & $SNR$ 17.7, $MAE$ 1.7 bpm\\
\cite{Mahmud2024} & CycleGAN  & wrist & $R_{HR}$  0.83 \\
\cite{Mohagheghian2024} & AE  & wrist & $SNR$ 4.4\\
\cite{AvilaCastro2025} & GAN  & finger & $SNR$ 11.8, $MAE$ 1.3 bpm\\
This study & CycleGAN & chest & $SNR$ 6.3, $MAE$ 2.0 bpm\\
\hline
\end{tabular}
\label{tab:literature}
\end{table}

\section{Conclusions}
The achieved findings highlights the potential of deep learning and style transfer techniques for improving wearable biosignal reliability, paving the way for accurate cardiac assessment and blood pressure estimation using chest-worn PPG sensors, in cases where finger sensors are impractical, such as during sleep or exercise. The proposed method ultimately enhances the reliability of wearable PPG-based monitoring, enabling accurate cardiovascular assessments without requiring extensive post-processing or hand-crafted noise models.

\section*{CRediT authorship contribution statement}
\noindent \textbf{Sara Maria Pagotto}: Conceptualization, Software, Data acquisition, curation and analysis, Writing – original draft. \textbf{Fedrico Tognoni}: Data acquisition and curation. \textbf{Matteo Rossi}: Core algorithm development. \textbf{Dario Bovio}: Acquisition system setup, Data curation. \textbf{Caterina Salito}: Acquisition system setup, Data curation. \textbf{Luca Mainardi}: Resources, Writing review. \textbf{Pietro Cerveri}: Conceptualization, Project administration, Funding acquisition, Supervision, Writing – review \& editing.

\section*{Declaration of interest statement}
\noindent The authors declare that they have no known competing financial interests or personal relationships that could have appeared to influence the work reported in this paper.


\section*{Acknowledgment}
\noindent This work was partially supported by Italian Ministry of Research and University - P.E. PE00000013-FUTURE ARTIFICIAL INTELLIGENCE RESEARCH (FAIR), and by EU Horizon-HLTH-20023-tool-05 “SMASH-HCM - Stratification, Management, and Guidance of Hypertrophic Cardiomyopathy Patients using Hybrid Digital Twin Solutions”, Grant 101137115.

\bibliographystyle{plain}

\end{document}